**Large array of sub-10 nm single-grain Au nanodots for use in nanotechnology**

*N. Clément[*], G. Patriarche[+], K. Smaali[*], F. Vaurette[*], K. Nishiguchi[#], D. Troadec[*], A. Fujiwara[#] and D. Vuillaume[*]*

[*] Institut d'Electronique Microélectronique et Nanotechnologie (IEMN), CNRS, Univ. of Lille, Avenue Poincaré, 59652, Villeneuve d'Ascq, France
E-mail: Nicolas.clement@iemn.univ-lille1.fr

[+] Laboratoire de Photonique et Nanostructures (LPN), CNRS, route de Nozay, 91460, Marcoussis, France

[#] NTT Basic Research Laboratories 3-1, Morinosato Wakamiya, Atsugi-shi, Japan

A uniform array of single-grain Au nanodots, as small as 5-8 nm, can be formed on silicon using e-beam lithography. The as-fabricated nanodots are amorphous, and thermal annealing converts them to pure Au single crystals covered with a thin $SiO_2$ layer. These findings are based on physical measurements, such as atomic force microscopy (AFM), atomic resolution scanning transmission electron microscopy, and chemical techniques using energy dispersive x-ray spectroscopy. A self-assembled organic monolayer is grafted on the nanodots and characterized chemically with nanometric lateral resolution. We use the extended uniform array of nanodots as a new test-bed for molecular electronics devices.

# 1. Introduction

Extensive and perfectly ordered arrays of close-to-identical nanoparticles (NPs) have attracted much attention for their numerous potential applications in several nanotechnology fields, including materials science[1,2], electronics[3,4], biology[5] and information technology [6,7]. Au nanoparticles a few nm across and distributed in a liquid solution are already commercially available. However, the poor controllability of their positioning and the difficulty of making reliable electrical contact limits their application. Therefore, several techniques to build arrays of NPs with well-defined dimensions and positions have been proposed recently[1,8-9]. A nanopost stamps technique[1] is one of the most promising methods in terms of small dimensions (~30 nm) and small dispersion.

Here we demonstrate the formation by e-beam lithography of sub-10 nm Au dots with small dispersion and perfect alignment. Such precise formation of small dots enables us to identify the critical size that determines whether a dot is composed of single or multiple crystal domains. Moreover, we show that annealing at moderate temperature can convert Au dots from amorphous to single crystal, and then they are covered with a thin $SiO_2$ layer. After easy removal of the $SiO_2$ (dilute HF etching), these nanodots can be used as electrodes for the chemical and electrical characterization of organic self-assembled monolayers (SAMs) with less than 200 molecules.

# 2. Array of sub-10 nm single-grain gold nanodots

Figs. 1a and 1b show scanning electron microscope (SEM) images of an array of Au nanodots fabricated on a highly doped n-type Si (100) substrate (see experimental section) with different zoom scales. The nanoparticle diameter can be estimated from the SEM images to be around 8 nm. This estimated size can be compared with that obtained by other methods: aberration-corrected scanning transmission electron microscopy (corrected-STEM) and atomic force microscopy (AFM), as shown in Figs. 1c and 1e. The STEM image shows sizes of 8 nm at the base and 5 nm at the top. The usual e-beam lithography approach for metal deposition after lift-off is the deposition of a bi-layer resist to form a "cap" after development. However, we were unable to reach dots smaller than 10 nm with this approach. The successful approach was to use a single diluted layer resist (see methods) as thin as possible (~40 nm). From the AFM image of the large array of nanodots (Fig. 1d), we obtained statistical distributions confirming that the nanodots are uniform, with height and diameter dispersions of 30 and 15 %, respectively. The average apparent dot radius obtained by AFM is overestimated due to the convolution effect of the tip. Considering[10] $D_{real} = D_{apparent} - 2(2R_{tip} \cdot h_{dot} - h_{dot}^2)^{1/2}$ where $D_{real}$ and $D_{apparent}$ are the estimated and measured dot diameters respectively, $R_{tip}$ the tip radius = 14 nm and $h_{dot}$ the dot thickness ~ 7.16 nm, we get $D_{real}$ = 9 nm, in agreement with other techniques. An example of an AFM image of a single dot, its cut view and deconvoluted profile are shown in Fig. 1f. Moreover, a Fourier transform of Fig. 1d shows the precise and well-ordered positioning of nanodots, with an average periodicity of 100 ± 1 nm, as shown in Fig. 1g.

**3. Critical size for single grain**

The size of nanodots can be controlled by adjusting the e-beam dose during the lithography process. A systematic study of doses and number of pixels allowed us to adjust the size from 5 to 50 nm as measured by SEM (see supplementary information, Fig. S1). We show that there is a critical size for nanodots that determines their crystalline structure. While smaller dots (~8 nm) are composed of a single domain (Fig. 2a), dots with a diameter of 25 nm are formed of multiple domains (Fig. 2b). The multiple domain structure can also be electrically confirmed with conducting AFM (C-AFM) images (Fig. 2c). In the multi-grain nanodots, dark features on the image and deep pitch in the current profile are easily observed by C-AFM at the grain boundaries. Single-grain nanodots appear very homogeneous. Thus the sub-10 nm dots are likely to be useful as nano electrodes in nanodevices, as shown below. The critical size for the monograin / multigrain transition is estimated from SEM images to be 15 nm (see supplementary information, Fig. S1). Theoretical estimation of such critical size is difficult due to the large number of parameters including nucleation points, surface diffusion constant, evaporation speed, nature and temperature of the substrate, film thickness and patterning of the substrate (e.g. lithography)[11]. It is likely that there is a competition between diffusion of gold atoms at surface and the density of nucleation points in silicon. For pattern larger than 15 nm, several nucleation points are involved and multi-grains are observed.

## 4. Thermal annealing: Crystal structure analysis and mechanism proposal

A cross-sectional view of a corrected STEM image (Fig.1c) shows that an Au nanodot is amorphous and that Au atoms diffuse into the silicon substrate. Since the image was taken just after the formation of the Au dots, the diffusion must have occurred during the metal evaporation step. The diffusion of an Au nanoparticle into silicon has been investigated by real-time (TEM) observation[12]. T. Ishida et al. have shown that the structure of the nanodots evolves with time from crystalline to amorphous as more Au atoms diffuse into silicon. This mechanism can explain the amorphous character observed in our nanodots.

From geometrical and electrical viewpoints, a crystalline structure is preferable to an amorphous structure. Thermal annealing can convert nanodots from amorphous to crystalline. Figs. 3a-c shows a cross-sectional STEM image of nanoparticles annealed at 260°C for two hours under a nitrogen atmosphere. We observe a crystalline structure with clear facets. Inter-planar lattice spacing evaluated from Fourier transform spectra of the atomic resolution STEM images (Fig. 3d) indicates a cubic structure with an evaluated lattice parameter of 0.407 nm, which is very close to the theoretical Au lattice parameter (0.40786 nm). This feature means that the crystallites shown by SEM images are single-crystal Au. More interestingly, the crystallographic orientation of the Au dots correlates with that of the silicon substrate: the <100> and <220> axes of the Au nanodots are aligned with the <100> and <220> orientations of the silicon substrate. Inverse Fourier transform images at a reflection angle specific to Au or silicon (Figs. 3d and 3e) indicate that pure Au crystallites penetrate into the silicon substrate. Figs. 3d and 3e correspond to dots with zone axes <100> and <110>, respectively. The micro twins that we sometimes observe inside the gold single crystal are Σ3 twins with {111} twin boundaries (Fig. 3f).

Chemical analysis with energy dispersive X-ray spectroscopy (EDX: see experimental section) of the area circled in red (Fig. 3b) indicates another important fact: the concentrations of Si and Au are 70 and 30%, respectively. As described above, the nanodots are composed of crystalline Au rather than the silicide Au-Si[6]. Therefore, EDX analysis indicates that the Au dots are covered with a thin $SiO_2$ layer. A detailed description of the protocol used to reach this conclusion is added in supplementary information with Fig. S2. $SiO_2$ cannot be directly observed by STEM due to the much smaller dimensions of the dots compared to the covering carbon thickness and the fact that the $SiO_2$ and carbon are both amorphous. We propose the following mechanism during annealing (Figs. 4a-d). First, Au atoms are deposited by evaporation and diffuse inside the silicon (Fig. 4a). A lift-off process forms amorphous Au nanodots as shown in Fig. 4b. An annealing process converts Au dots from amorphous to crystalline. Simultaneously, Si atoms coexisting with Au atoms migrate to the surface of the Au nanodots (Fig. 4c). This is in agreement with previous studies of thin gold films evaporated on silicon and annealed at temperatures below the eutectic point (370°C).[13] From the extent of the reduced Si area shown by Fig. 3c, one can estimate that a layer of Si atoms 3 nm thick covers each nanodot. This Si layer is then oxidized to form $SiO_2$ (~ 5 nm) when exposed to ambient conditions (Fig. 4d). In some cases, during the annealing process when Si atoms diffuse at the surface, some Au atoms are carried out with Si and one can observe a second Au nanoparticle lying on top of the $SiO_2$ that covers a primary Au nanoparticle (Fig. 4e).

We have performed a systematic study of several dot sizes, annealing times, and temperatures. We observed different consequences of annealing depending on dot size. When small dots (e.g. 8 nm) were annealed at 200 °C for two hours (Fig. 5a), their shape remained similar to that of unannealed ones (Fig. 1c). Since their structure was crystalline (with macles) as seen in the STEM image in Fig. 5a, but facets were not observed as in single-grain nanodots (Fig. 3c), this case can be considered to be an intermediate step between amorphous dots and dots annealed at 260°C. Another interesting phenomenon is related to diffusion in the 10 nm range. For large dots (e.g. 60 nm), which appear to be multi-grain structures (Fig. 5b), grains merge after thermal annealing to form a more compact structure. An SEM image of such dots annealed at 300°C for 20 minutes is shown in Fig. 5c. The dot diameter is reduced by about 20 nm after thermal annealing, with dark areas observable around the dots that are related to zones from which gold has migrated (see the STEM images in Figs. 5c and 5e). These merged dots (e.g. annealed at 300 °C) are crystalline, while dots annealed at lower temperatures (e.g. 150 °C) are polycrystalline (Fig. 5d). Fig. 5d summarizes the critical temperature $T_c$ (annealing for 1 h) and critical time $t_c$ (annealing at 300 °C) corresponding to the transition from multi-grain to polycrystalline dots, as obtained by direct observation of SEM images (see supplementary information Figs. S3 to S8). Since thermal annealing temperatures have been tested every 50 °C and annealing times every 10 min, error bars are half of these values. While $t_c$ barely varies with dot diameter, $T_c$ increases. $T_c$ is likely related to the diffusion length of gold atoms. $t_c$ was measured for an annealing temperature (300 °C) above $T_c$ (1 h annealing) for all dot sizes, which could explain the small dependence of $t_c$ on dot size. For application purposes, this graph shows that annealing at 300 °C for 20 min is enough to merge the dots. For molecular electronics applications, we will focus on the smallest dots, because the diffusion process often induces the presence of SiOx between some gold atoms attached to silicon and the gold dot, an effect clearly seen in STEM images of 15 nm and 30 nm dots (Fig. S9).

**5. Metal/Molecules/Metal junctions with gold nanodots as electrodes**

The oxide layer covering an Au dot is easily removed by using dilute HF solution, and since the bottom of the Au dot forms a good ohmic contact with the highly doped Si substrate, these nanodots are very interesting for the fabrication of a metal-molecule-metal (MMM) nano-junction as shown in the inset of Fig. 5a. We formed a self-assembled monolayer (SAM) of alkylthiol molecules using well documented recipes[14] ($C_{12}H_{25}$-SH) on the 8-nm diameter nanodots, and we measured current-voltage characteristics as shown in Fig. 5a by contacting the SAM with the tip of a C-AFM. The number of molecules between one Au nanodot and the tip of the C-AFM is estimated to be 80, with the assumption of a molecule coverage of about 25 Å² per molecule[16]. Without molecules, i.e. junctions composed of a direct Au dot / C-AFM (Pt) tip contact showed ohmic current characteristics (Fig.5a, inset). The current characteristics of the nanodot/molecule/C-AFM tip nanojunctions show a lower conductance. These I-V curves are dominated by tunneling behavior, as usually observed in the MMM junctions of alkyl chains[20]. Considering the Simmons model for a tunnel current through a rectangular barrier[15] without image charge consideration, a barrier thickness of 17 Å, an effective mass of 0.2, a contact area of 48 nm² and tunnel barrier heights of 1.5 eV and 1.6 eV for positive and negative voltages respectively, give reasonable fits as shown in Fig. 5a. Minima voltages obtained by transient voltage spectroscopy[22] (Fig. 5b) are -1.4 V and +1.3 V, in agreement with tunnel barrier heights used to fit I-V curves and published data for alkyl chains[20]. The small asymmetry in I-V curves at positive and negative voltages can be explained by the small difference in work function between the CAFM tip (Pt) and Au nanodot. These results show that these Au single-strain nanodots tend to make good molecular junctions. Moreover, we can analyze current characteristics statistically over thousands of junctions, using only a single C-AFM image, and this approach gives very significant and valuable statistical data on the electronic transport properties of molecular devices, as we report elsewhere[16].

The MMM using nanodots with a sub-nm lateral size provides a new test-bed for molecular junctions to close the experimental gap between single-molecule experiments (STM and beak-junction)[17] and nanopore junctions (size of about 30 nm)[18]. This approach also confers the ability to easily fabricate a single-domain electrode made of single-crystalline Au, instead of the polycrystalline one typically used for C-AFM experiments on large metal surfaces[19-20].

## 6. Gold nanodots on SiO$_2$ / Chemical analysis of ferrocene molecules on nanodots with nanometric lateral resolution.

As well as on doped Si, Au nanodots can also be formed on an SiO$_2$ substrate. Au dots on an SiO$_2$ substrate are useful for optical applications that require an insulating layer below the dots (e.g. fluorescence spectroscopy), as well as for sensors using a floating gate[21]. Chemical analysis of the molecules on nanodots is of prime importance because, by doing so, we can expect to establish a relationship between what has been measured (current, fluorescence, etc...) and the chemical nature of the SAM on these nanodots. The inset of Fig. 6 shows a cross-sectional STEM view of Au nanodots on 90-nm-thick thermally-grown SiO$_2$. The dots are covered (see supplementary information) by a SAM of ferrocenyl hexane thiol molecules (FeC$_6$H$_{21}$-SH), which is used as an archetype for several applications (e.g. molecular memories[22], ...). The curved shape of the dots after annealing is different from dots on a Si substrate because of the amorphous structure of the SiO$_2$ below the dots. Fig. 6 is a high-angle annular dark field (HAADF-STEM) image of one dot. Fe atoms can be distinguished on this image. The HAADF-STEM is also referred as "Z contrast" imaging, and single atoms with a high atomic number give a bright spot on the HAADF image. EDX analysis at five different zones (see Fig. 6) is listed in Table 1. For the EDX quantification analysis, we used K-lines for S and Fe and L-lines for Au. The concentrations of Fe and S indicate that ferrocene molecules are only located on the nanodot. We have observed no quantum effects related to the nanoparticle itself. The average spacing of successive quantum levels $\delta$ (Kubo gap) remains smaller than kT at room temperature for gold nanoparticles > 2-3 nm[23]. Therefore, our nanoparticles are expected to be metallic.

## 7. Conclusions

To conclude, we have demonstrated the fabrication of a precisely aligned large array of sub-10 nm single-grain Au nanodots on Si and $SiO_2$. When lithography-defined nanodot dimensions exceed 15 nm, multiple grains of Au are observed in each dot. Below 15 nm, an as-fabricated amorphous nanodot evolves to a single-crystal nanodot when annealed at 260°C. For larger dots, a polycrystal is formed by annealing at a critical temperature that depends on dot diameter. At this stage, Si atoms that were coexisting with Au atoms diffuse at the nanodot surface and form ultrathin silicon oxide. Nanodots on a conductive Si substrate were used as an extremely small electrode to measure the electrical properties of a molecular junction made of less than a hundred organic molecules. We also demonstrate that chemical analysis at the nanoscale can be used to characterize these molecularly functionalized nanodots. Since Au nanodots fabricated using the e-beam lithography and annealing process have high geometrical, positional and crystal uniformity, they are very useful for any study that requires single-crystal Au with a good electrical contact.

## 8. Experimental

*Nanodot fabrication:*

On Si: For e-beam lithography, we use an EBPG 5000 Plus from Vistec Lithography. The (100) Si substrate (resistivity = $10^{-3}$ Ω.cm) is cleaned with UV-ozone and native oxide etched before resist deposition. The e-beam lithography dimensions have been optimized by using a 45 nm-thick diluted (3:5 with anisole) PMMA (950 K). For the writing, we use an acceleration voltage of 100 keV, which reduces proximity effects around the dots, compared to lower voltages. We tried different beam currents to expose the nanodots (100 pA and 1 nA), and we could see no difference in the size of the nanodots as a function of current. So, for the final process, we used 1 nA to optimize exposure time. Then, the conventional resist development / Au – e-beam evaporation (8 nm) / lift-off processes are used. Immediately before evaporation, native oxide is removed with dilute HF solution to allow good electrical contact with the substrate.

On $SiO_2$, the same resist is used as for the Si substrate. 1 nm of Ti is deposited by e-beam evaporation before Au evaporation to ensure good adherence of the dots on the $SiO_2$ surface.

*Sample preparation for TEM analysis*

Amorphous carbon is deposited by thermal evaporation on the full surface. The advantage of carbon is that it is amorphous and transparent to electrons. Then Pt is deposited by electron beam ( thickness ~ 50 nm) and locally above the nanodots by focused ion beam (FIB) (thickness ~ 500 nm). This two-step Pt deposition has the advantage of reducing the risk of the layer of interest being degraded by Ga ions. The foil is estimated to be less than 100 nm thick.

*TEM analysis*

We used a Jeol 2200FS microscope, equipped with a spherical aberration corrector on the probe for the STEM mode, working at an accelerating voltage of 200 keV. For the STEM images, as well as for the EDX analysis spots, we use a convergent probe of half-angle 30 mrad and a current of 150 pA, allowing a spot size of 0.13 nm. The inner and outer annular angles of the HAADF detector were 100 and 170 mrad, respectively.

*EDX resolution*

The probe used for EDX is the same as the one used for atomic resolution images. At the cross-over point (inside the dot), the diameter is 0.1 nm. The analyzed volume is related to the probe angle (half-angle = 30 mrad). We have therefore a volume composed of 2 cones. Given a sample thickness of 100 nm (typical value for FIB preparation), we estimate the maximum diameter to be 1.5 nm. The probe current used was 150 pA.

*EDX analysis of ferrocene molecules*

We used the K-lines for S and Fe and the L lines for Au. As a thick layer of amorphous carbon covers the dots, this light element is detected but not taken into account for the quantification. The M-lines of Au ($M_\alpha$ : 2.123 keV, $M_\beta$ : 2.205 keV, $M_{a,b}$: 2.307 keV) are in the same energy position as the K-lines of S (K : 2.308 keV), hence it is impossible to measure low concentrations of sulfur associated with gold.

**Acknowledgements**

The authors would like to thank C. Boyaval from IEMN for assistance in SEM imaging.

**Supporting information**

Supporting information is available online

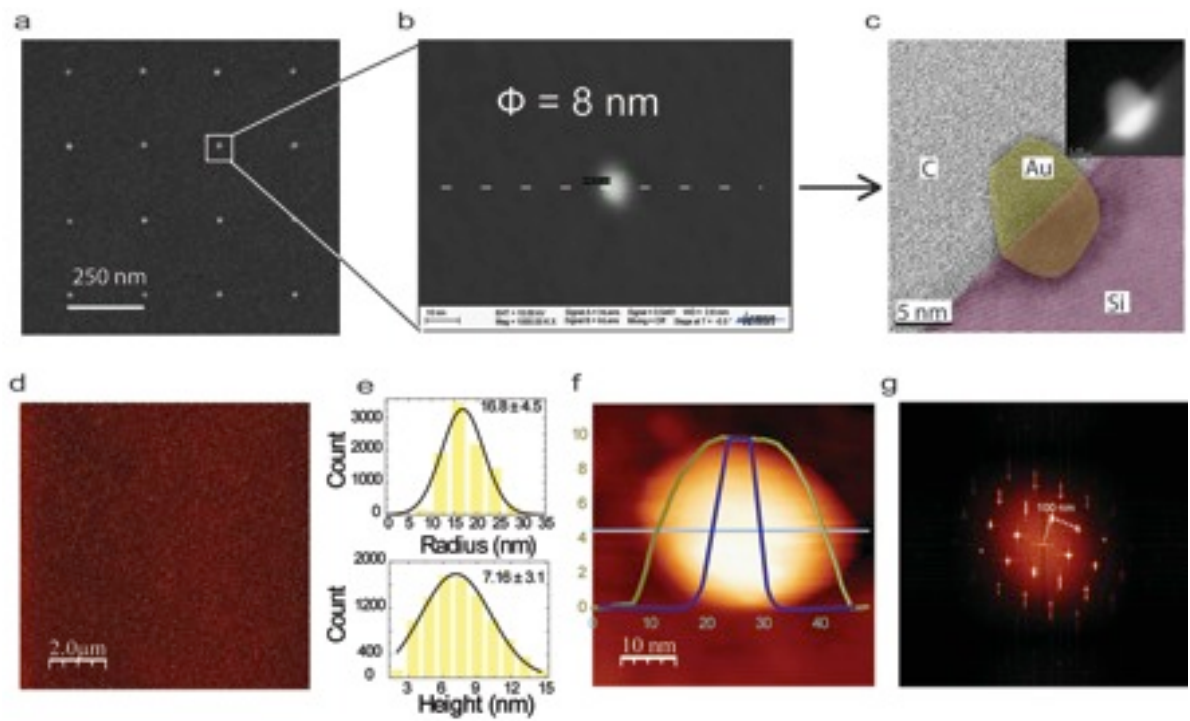

**Figure 1.** a-b) SEM (10 kV) images at different zoom scales of an array of gold nanodots. c) colored HRTEM image of a single dot (inset is the raw image) d) AFM image obtained with a sharp tapping-mode tip (radius ~ 15 nm) from Nanosensors ©. e) histograms from 9500 samples (height and radius) of our nanodots. Dots radius are overestimated due to tip convolution. f) AFM image of a single dot with a superimposed profile along the grey line. g) Fourier transform image of the array shown in Fig. 1d.

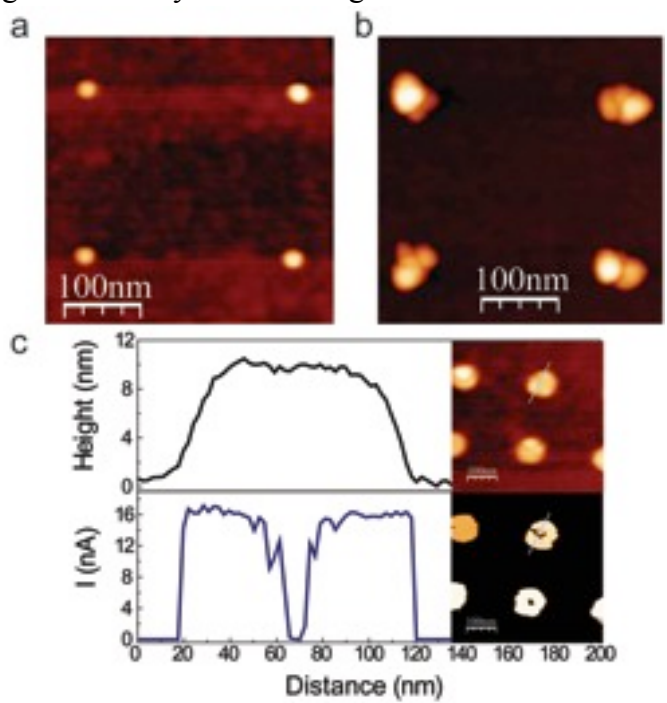

**Figure 2.** a) AFM image of four single-grain gold nanodots. b) AFM image of four multi-grain gold nanodots. c) AFM and CAFM images (tip voltage = 10 mV) of a large dot with cut views along the grey lines.

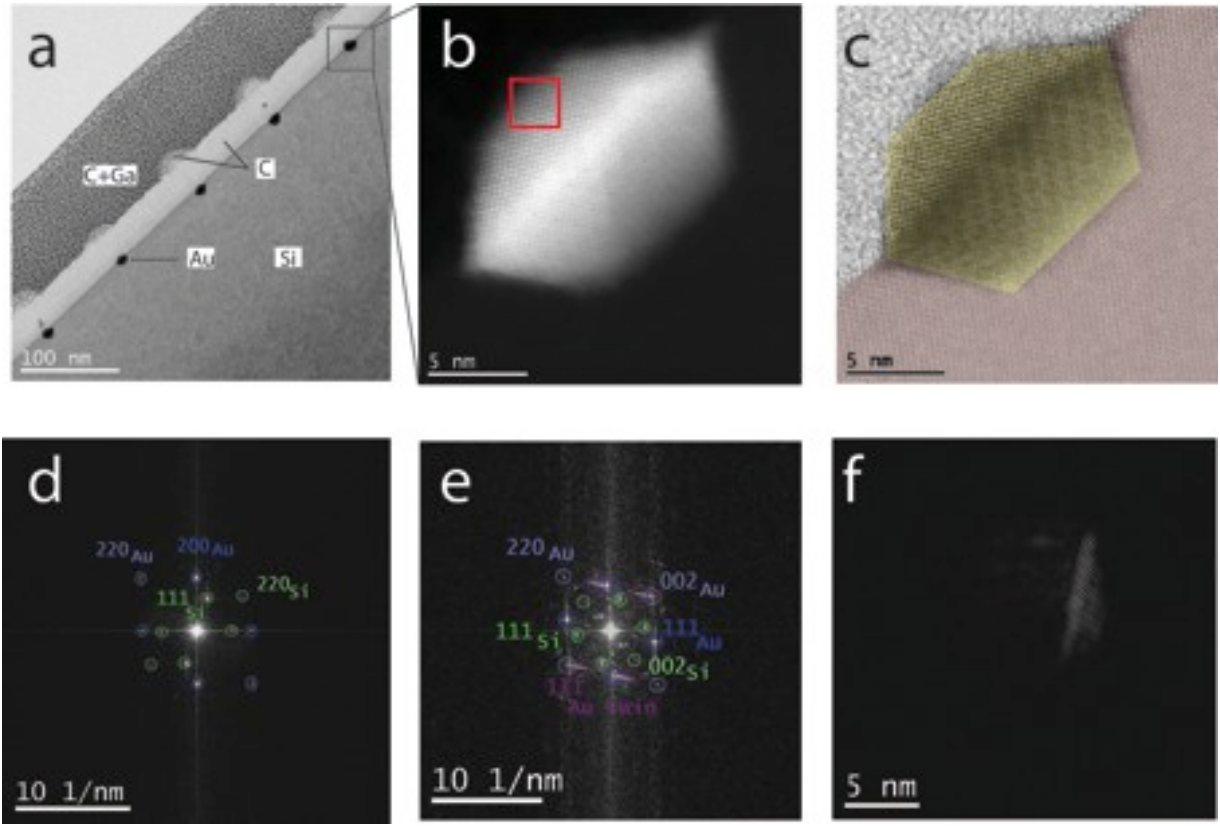

**Figure 3.** a) STEM image showing the bulk Si, five annealed dots, carbon layer and Pt layers. b) STEM image of a single annealed nanodot (260°C-2h). c) colored image of b). d) Fourier Transform image of a nanodot with zone axis <100> and e) Fourier transform of a dot with zone axis <110>. f) Inverse Fourier transform image corresponding to e)

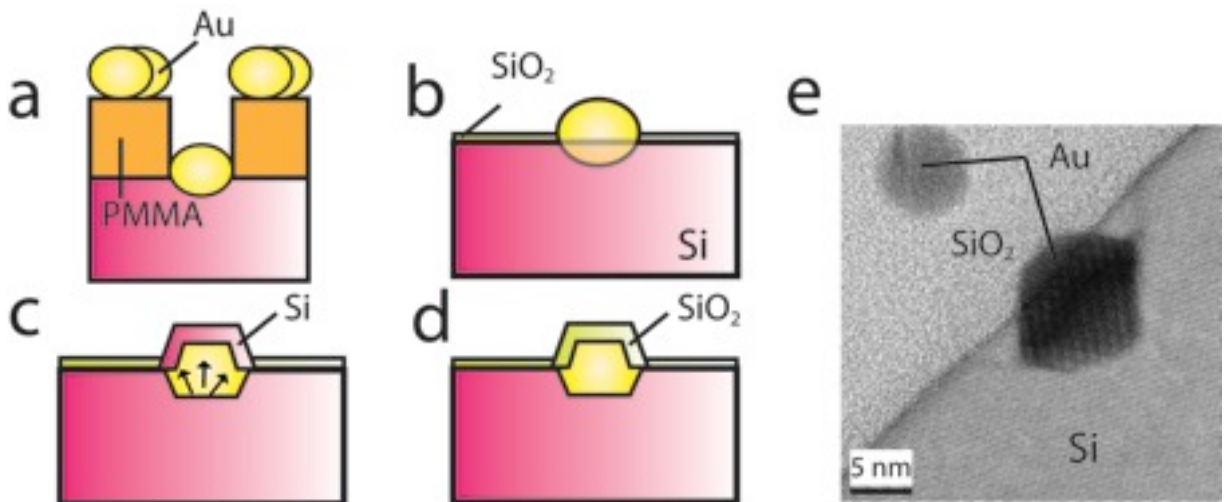

**Figure 4.** a-d) The fabrication of monocrystal gold nanodots. a) evaporation of 8 nm of gold after e-beam lithography and HF etching. b) lift-off process. The green surface covering Si represents the native silicon oxide. c) during thermal annealing, silicon diffuses at the surface d) in air, silicon covering the gold becomes oxidized. e) atomic resolution STEM image of a gold nano-dot with an extra gold particle covered by $SiO_2$.

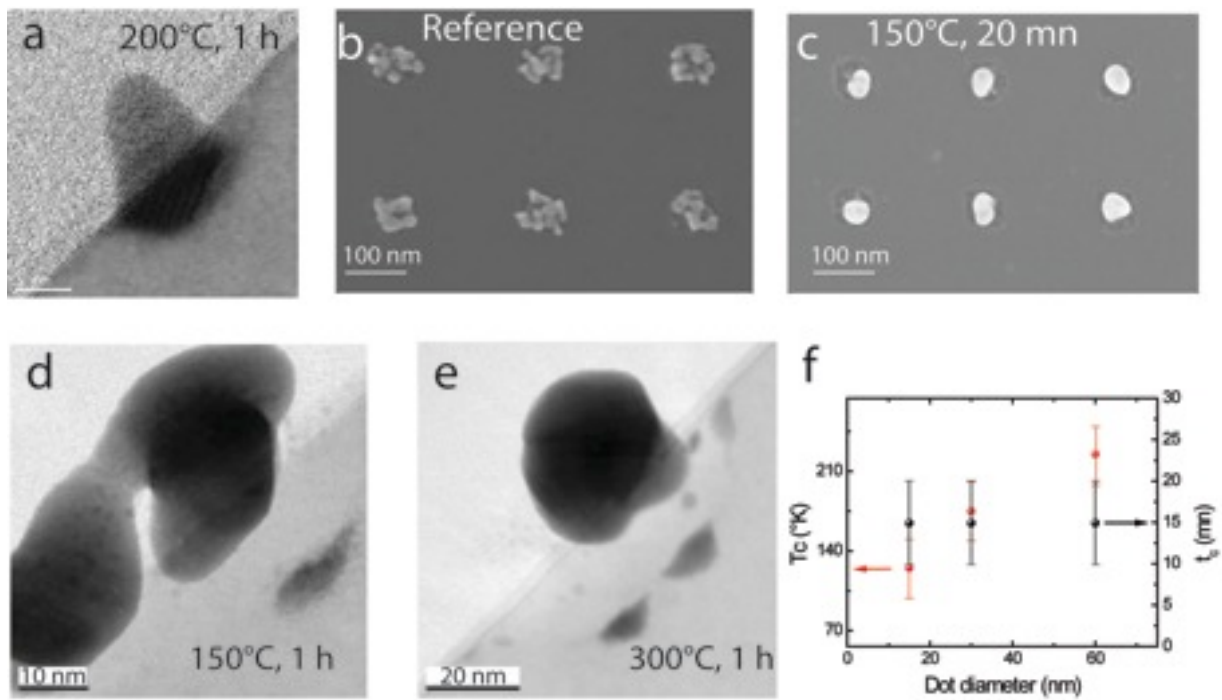

**Figure 5.** a) STEM image of a single-grain Au nanodot annealed at 200°C for 2 hours. b) SEM image of 60 nm wide multi-grain Au dots without annealing. c) SEM image corresponding to b) after annealing at 150°C for 20 min. d) STEM image of multi-grains after annealing at 150°C for 1h. e) STEM image of a merged gold dot after annealing at 300°C for 1h. d) Critical transition temperature $T_c$ (annealing during 1h) and time $t_c$ (annealing at 300°C) corresponding to the transition from multi-grain to poly-crystal as a function of dot diameter. This graph is obtained by direct observation of the SEM images shown in supplementary information (Figs. S3-S8).

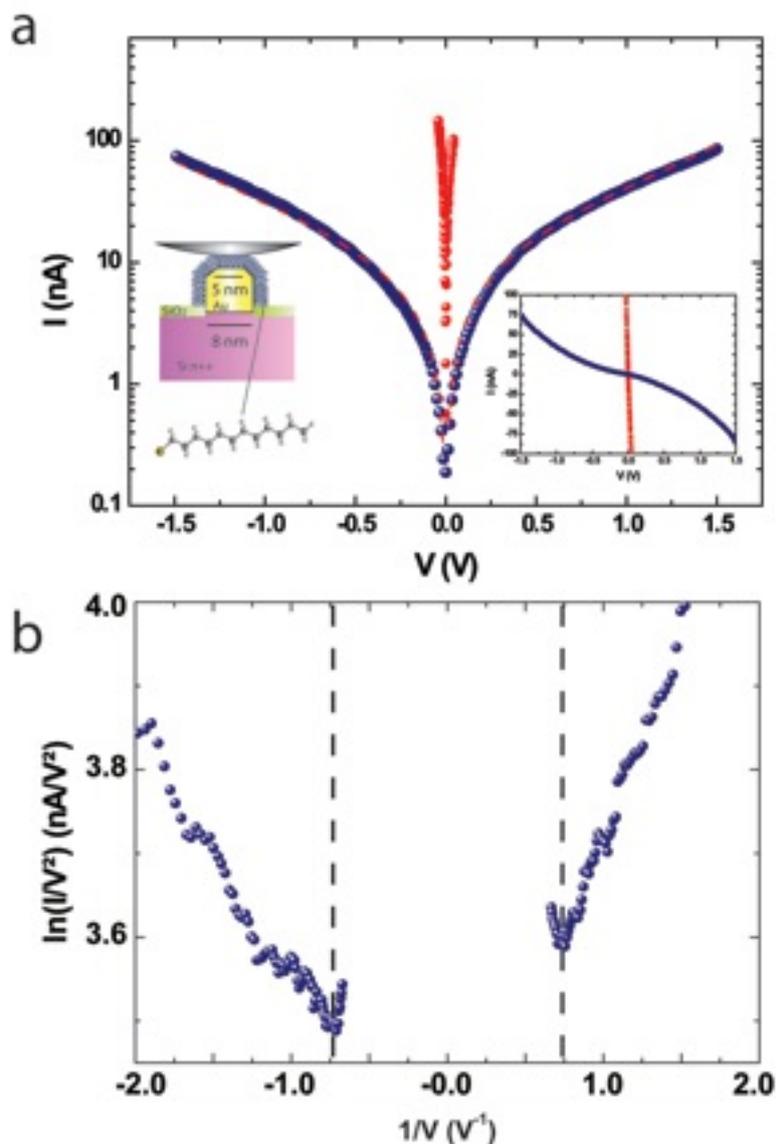

**Figure 6.** a) Inset, left: schematic view of a molecular junction using a gold nanodot as a bottom electrode and a CAFM PtIr (radius ~ 30 nm, force ~ 18 nN) tip as an upper electrode. We have covered nanodots with $C_{12}H_{25}$-SH alkyl chains. The current-voltage characteristics of the reference (without molecules) and of the molecular junction are shown. Fits using Simmons equation are shown as dashed lines. Inset right, same graph shown with linear scale. b) Transient voltage spectroscopy corresponding to curve a). Minima are indicated by dashed lines.

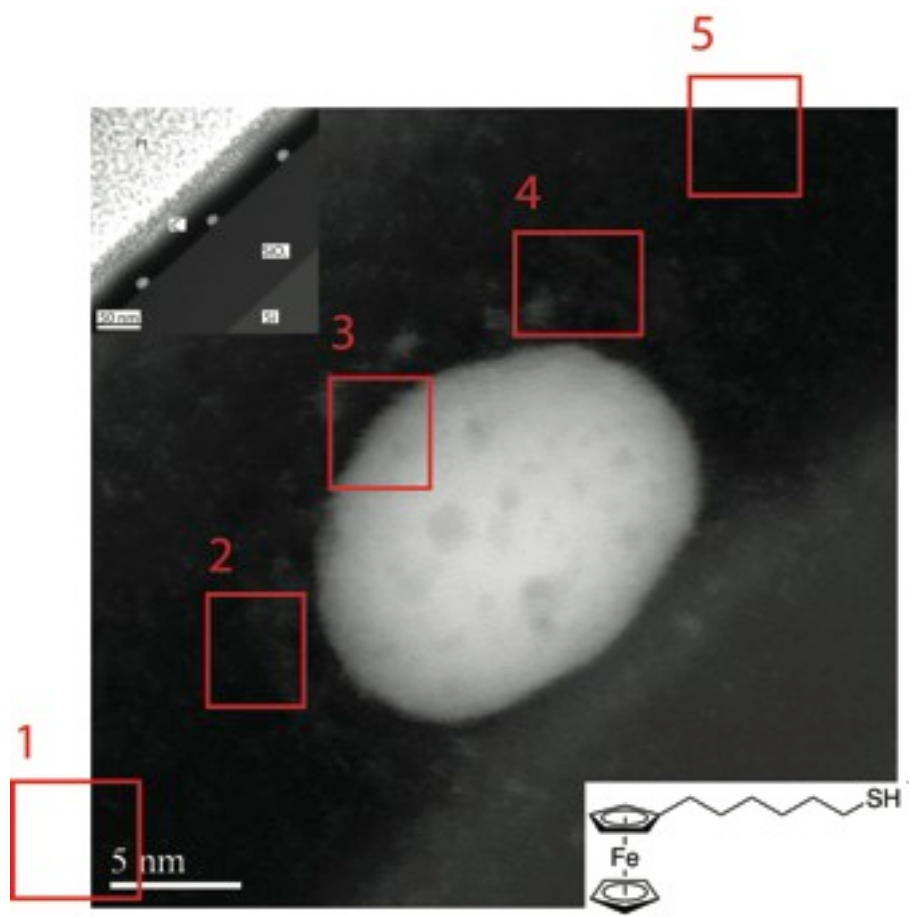

**Figure 7.** Inset: STEM image of 3 gold nanodots on SiO$_2$ (90 nm) covered by ferrocenyl hexane thiol molecules. Main figure: HAADF image of a single dot. White areas on the dot show the presence of Fe atoms with nanometric resolution. The different zones selected for EDX analysis (reported in Table 1) are numbered and shown with red boxes.

**Table 1.** Concentrations of Fe, Au and S obtained by EDX at the five different zones marked in Fig. 6.

| Zone # | 1 | 2 | 3 | 4 | 5 |
|---|---|---|---|---|---|
| Fe (%) | 0 | 0.47 | 0.68 | 0.5 | 0 |
| Au (%) | 0 | 0.01 | 0.2 | 0.03 | 0.03 |
| S (%) | 0 | 0.23 | Not detected | 0.18 | 0 |

**Supporting information**

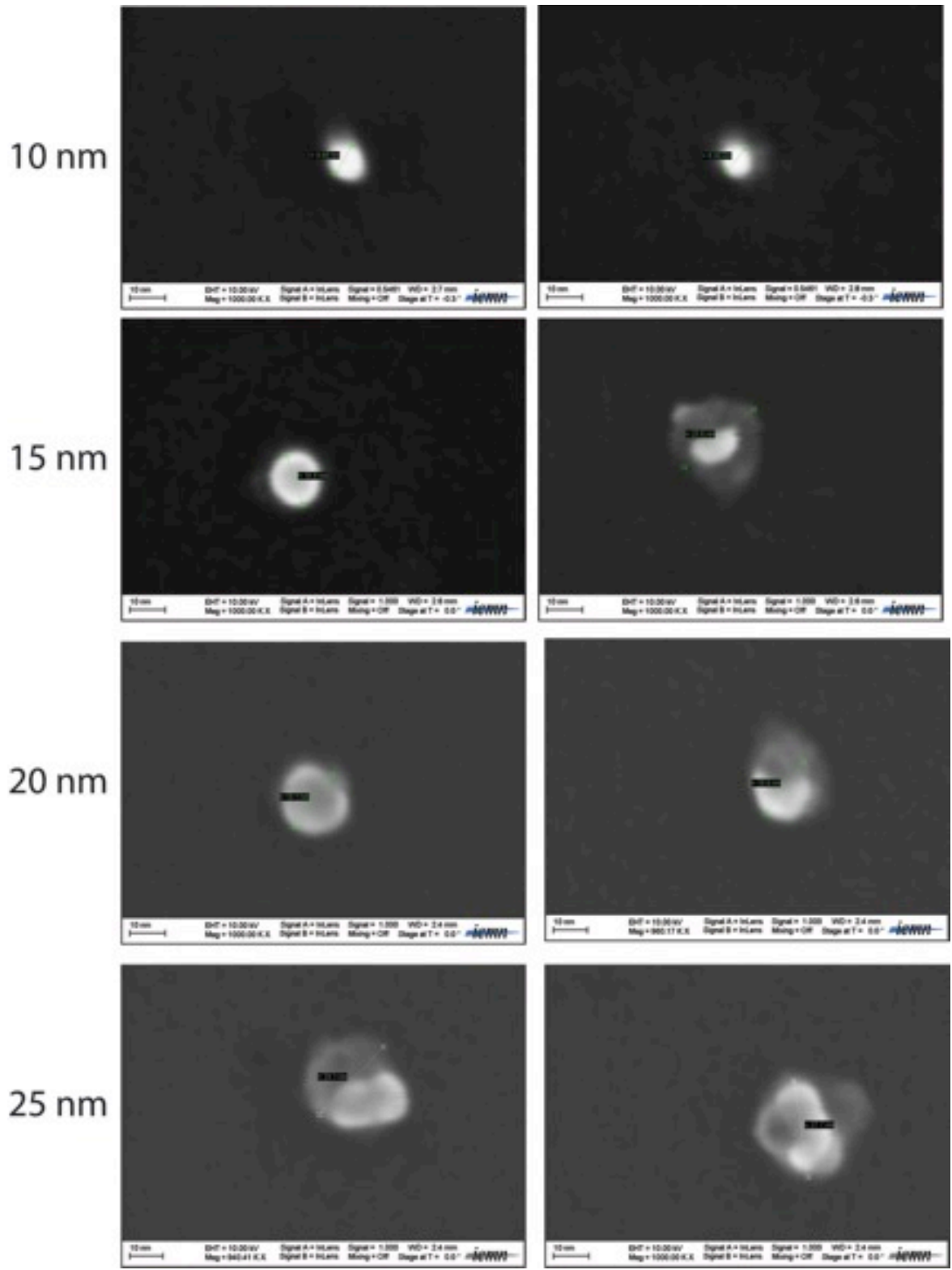

**Fig. S1** SEM images showing the evolution from monograin to multigrain structure as dot size increases. From 20 nm, we systematically observe multigrains.

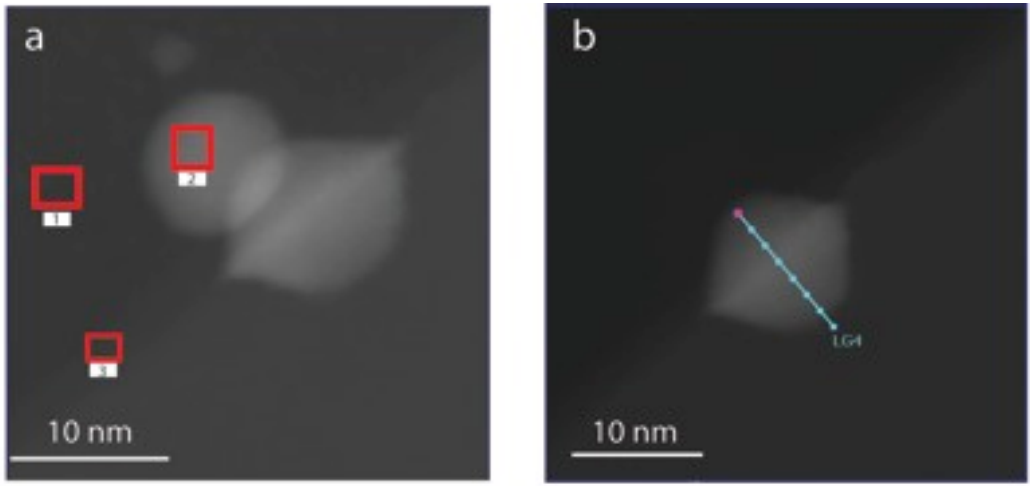

**Fig. S2** a) HRTEM image of a gold nanodot linked to the Si substrate and a gold particle lying on top. EDX analysis was performed in 3 zones (marked in red). Zone1 is the reference in the carbon layer (0.4% Si, 1.5% O, 0% gold, 98% C ). Zone 2 is on the gold particle: 3% Au, 3% Si, 8% O, 86% C. The level of oxygen is far larger than in the reference. Since the particle is composed of pure gold (obtained from interatomic distances), the other elements are around the particle. The ratio Si/Au is ~1 therefore we find SiOx with x close to 1.5 around the particle. Zone 3 is the reference of native oxide SiOx with x~1.3.

b) STEM image of a gold nanodot. EDX analysis was performed at different positions along the blue line.

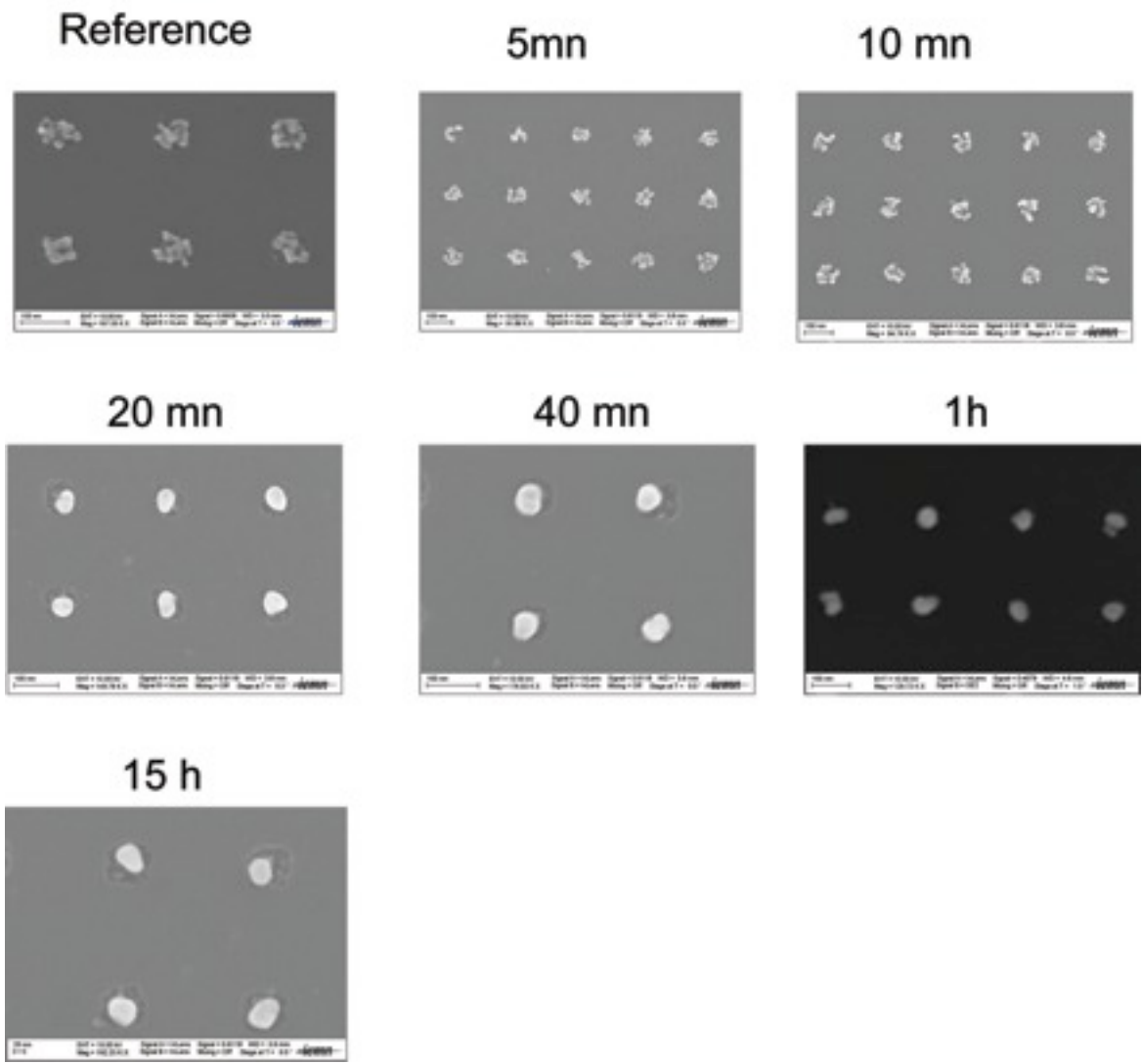

**Fig. S3** SEM images showing the effect of annealing time (at 300 °C) for large dots (e.g. 60 nm).

## Diameter 30 nm

Reference | 5 mn | 10 mn

20 mn | 40 mn | 1 h

15 h

**Fig. S4** SEM images showing the effect of annealing time (at 300 °C) for intermediate size dots (e.g. 30 nm).

## Diameter 15 nm

**Reference** | **5 mn** | **10 mn**

**20 mn** | **40 mn** | **60 mn**

**15 h**

**Fig. S5** SEM images showing the effect of annealing time (at 300 °C) for small dots (e.g. 15 nm).

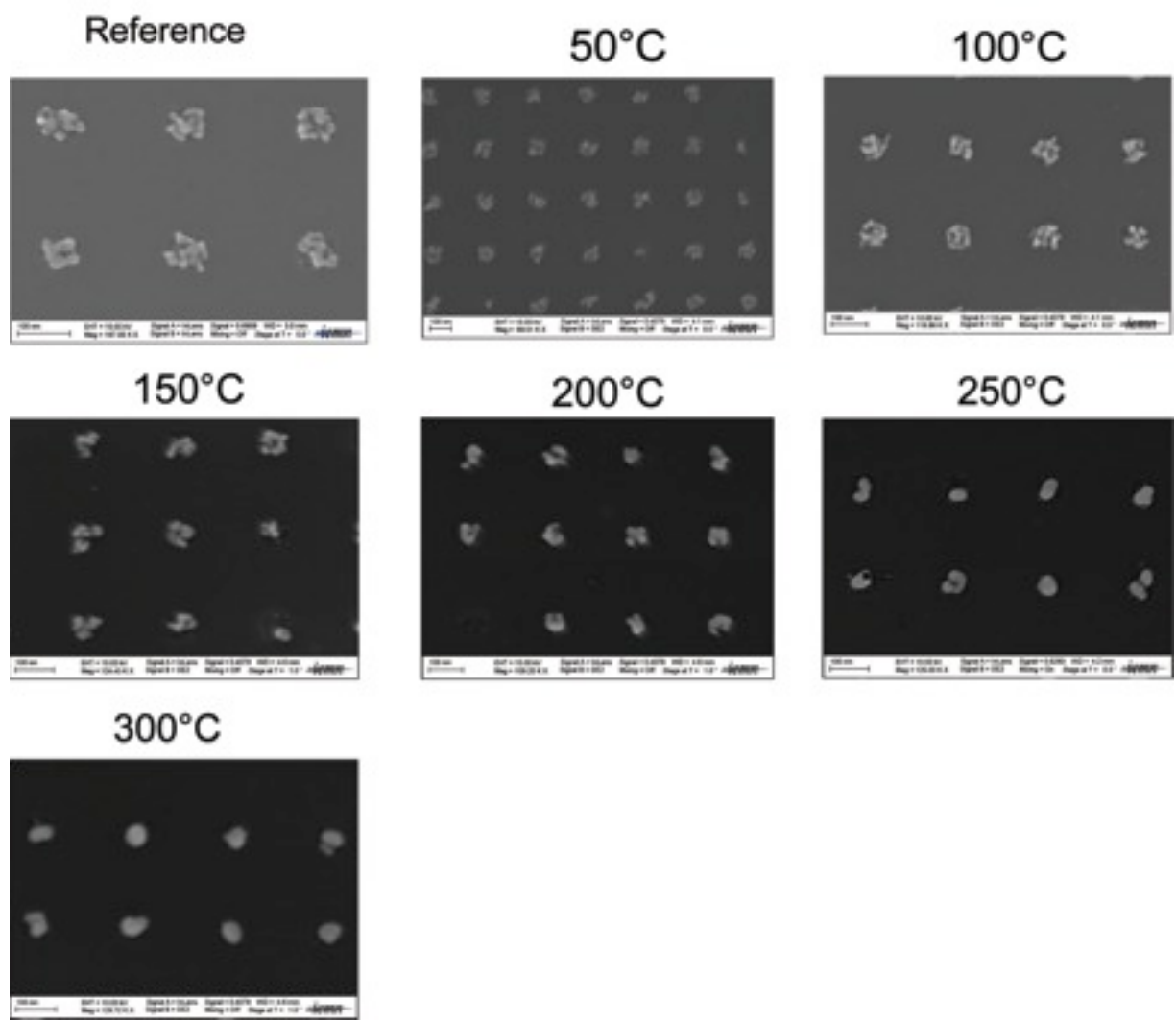

**Fig. S6** SEM images showing the effect of annealing temperature (for 1 h) for large dots (e.g. 60 nm).

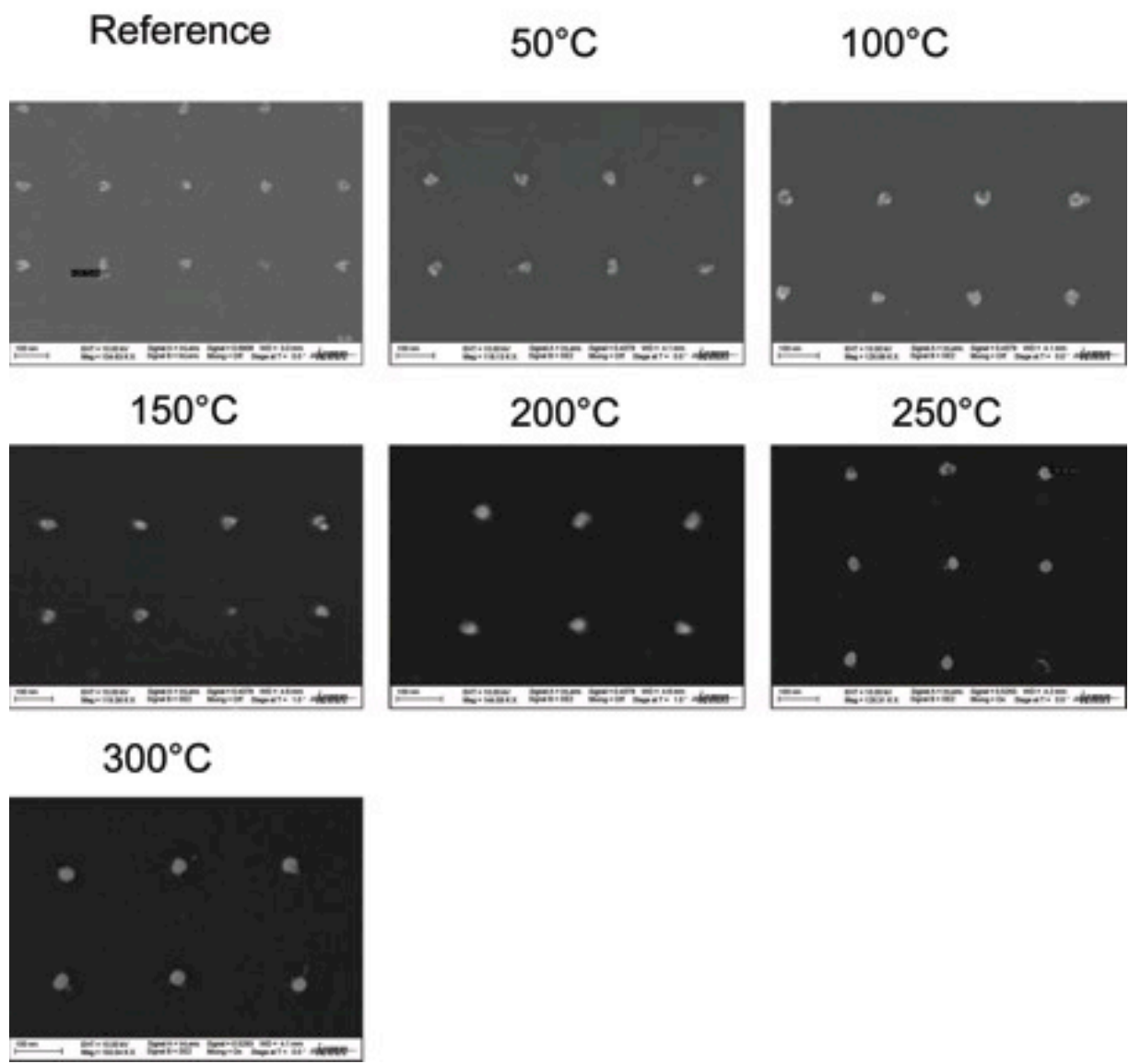

**Fig. S7** SEM images showing the effect of annealing temperature (for 1 h) for intermediate-size dots (e.g. 30 nm).

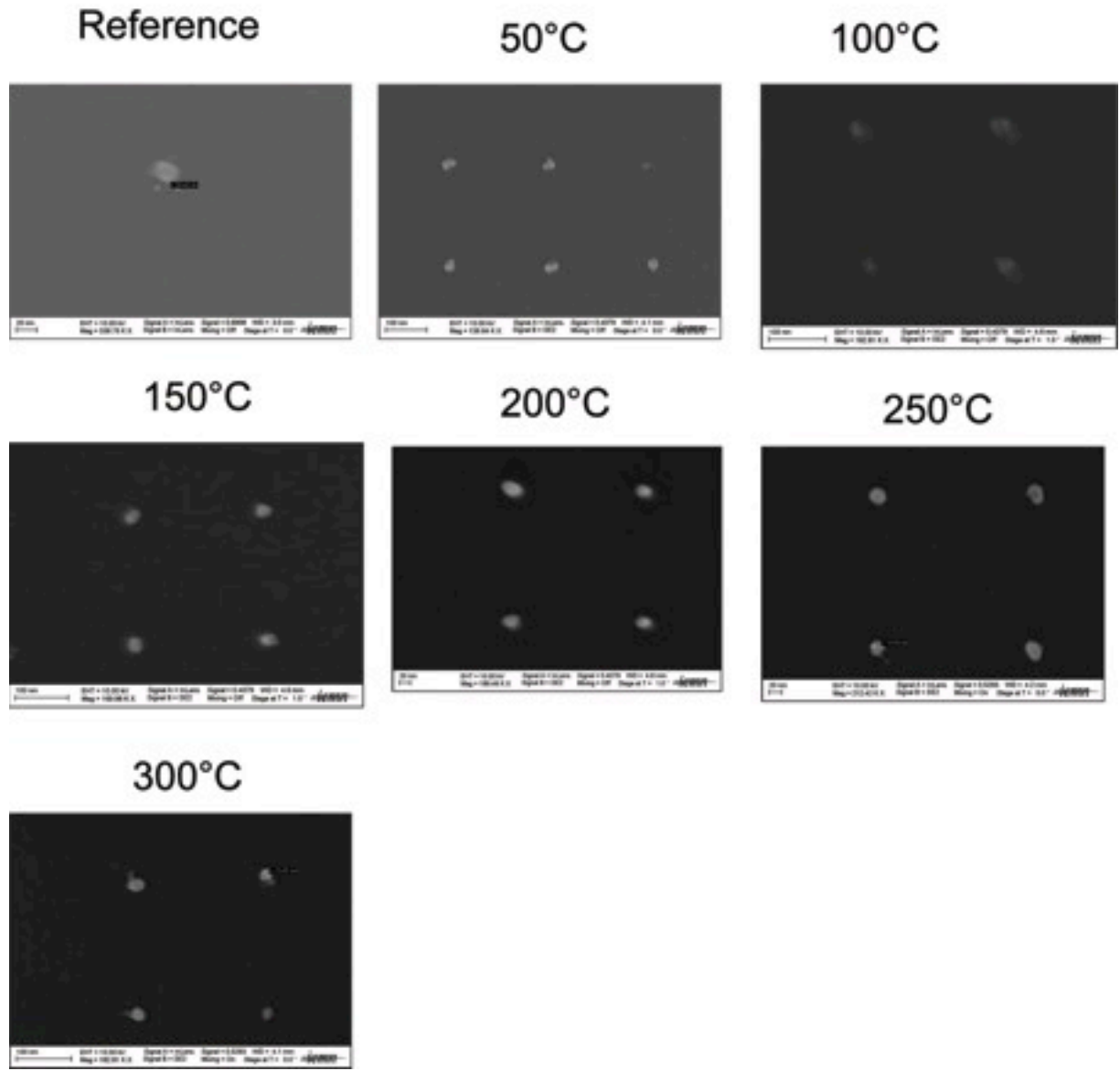

**Fig. S8** SEM images showing the effect of annealing temperature (for 1 h) for small dots (e.g. 15 nm).

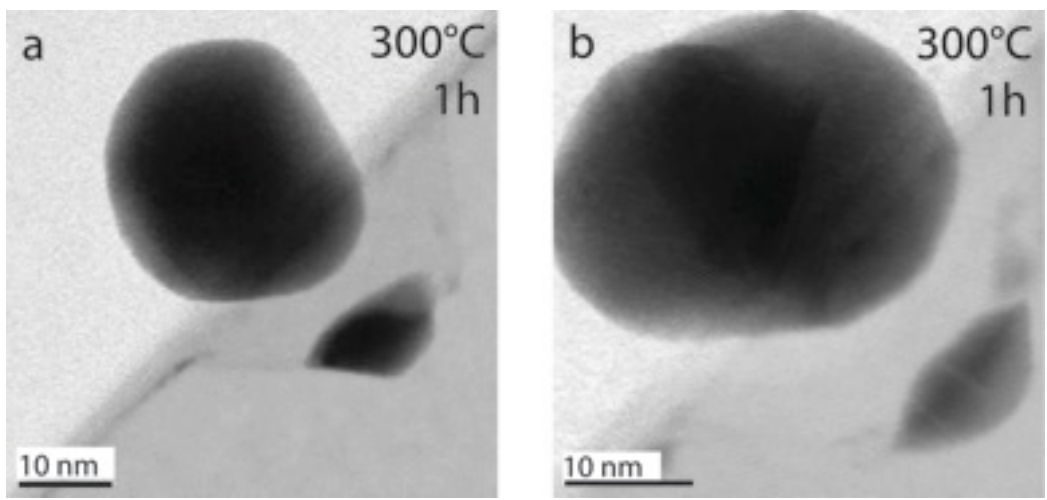

**Fig. S9** HRTEM images of 15 nm wide a) and 30 nm wide b) gold nanodots annealed at 300°C for 1 h.